\documentclass[conference]{IEEEtran}
\IEEEoverridecommandlockouts

\usepackage[style=ieee,minnames=1,maxcitenames=2]{biblatex} 
\usepackage{amsmath,amssymb,amsfonts}
\usepackage{algorithm}
\usepackage{algpseudocode}
\usepackage{graphicx}
\usepackage{textcomp}
\usepackage{xcolor}
\usepackage{subcaption}
\usepackage[inline]{enumitem}
\usepackage{balance}

\begin{document}

\title{Bugsplainer: Leveraging Code Structures to Explain Software Bugs with Neural Machine Translation}

\author{
\IEEEauthorblockN{Parvez Mahbub}
\IEEEauthorblockA{
\textit{Dalhousie University, Canada}\\
parvezmrobin@dal.ca}
\and
\IEEEauthorblockN{Mohammad Masudur Rahman}
\IEEEauthorblockA{
\textit{Dalhousie University, Canada}\\
masud.rahman@dal.ca}
\and
\IEEEauthorblockN{Ohiduzzaman Shuvo}
\IEEEauthorblockA{
\textit{Dalhousie University, Canada}\\
oh599627@dal.ca}
\and 
\IEEEauthorblockN{Avinash Gopal}
\IEEEauthorblockA{
\textit{Metabob Inc., USA}\\
avi@metabob.com}
}

\maketitle

\begin{abstract}
Software bugs cost the global economy billions of dollars each year and take up $\approx$ 50\% of the development time.
Once a bug is reported, the assigned developer attempts to identify and understand the source code responsible for the bug and then corrects the code. 
Over the last five decades, there has been significant research on automatically finding or correcting software bugs. However, there has been little research on automatically explaining the bugs to the developers, which is essential but a highly challenging task.
In this paper, we propose Bugsplainer, a novel web-based debugging solution, that generates natural language explanations for software bugs by learning from a large corpus of bug-fix commits. Bugsplainer leverages code structures to reason about a bug and employs the fine-tuned version of a text generation model -- CodeT5 -- to generate the explanations. \\
Tool video: https://youtu.be/xga-ScvULpk

\end{abstract}


\section{Introduction}
A software bug is an incorrect step, process, or data definition in a computer program that prevents the program from working correctly~\parencite{ieee-standard}.
Each year software bugs cost the global economy billions of dollars~\parencite{zou2018practitioners}.
Therefore, fixing bugs effectively is essential for both software practitioners and users. 
According to several studies, bug-fixing takes up $\approx$~50\% of the development time~\parencite{Britton2013} and consumes up to 40\% of the total budget~\parencite{Glass2001}.
Over the last five decades, there have been numerous studies to automatically find the location of bugs~\parencite{zou2018practitioners, rahman2021forgotten} and to automatically correct the buggy code~\parencite{jiang2021cure, li2020dlfix}.
However, neither many studies attempt to explain the bugs in the source code to the developers, nor are they practical and scalable enough for industry-wide use~\parencite{kochhar2016practitioners, zou2018practitioners}.

Explaining a bug in source code is necessary for understanding and fixing the bug, but a difficult task.
Many existing tools such as FindBugs~\parencite{FindBugs}, PMD, SonarLint, PyLint, and pyflakes~\parencite{pyflakes} employ complex hand-crafted rules and static analysis to detect bugs and vulnerabilities in source code.
Unfortunately, the value of the explanations 
from traditional static analysis tools may be limited due to 
their high false-positive results and the lack of actionable insights~\parencite{barik2016should, johnson2013don, ayewah2007evaluating}.
Unlike traditional rule-based approaches, explaining software bugs can be viewed as a machine translation task, where the buggy code is the source language, and the corresponding explanation is the target language.
In recent years, machine translation, especially neural machine translation (NMT)~\cite{jurafsky_martin_2014}, has found numerous applications in different software engineering tasks including, but not limited to, code summarization~\cite{loyola2017neural, hu2018deep}, code comment generation~\cite{tufano2021towards, hong2022commentfinder}, and commit message generation~\cite{jiang2017automatically, tao2021evaluation, liu2020atom, liu2018neural}.
Recent advancements in large language models (LLM) have led to general-purpose tools (e.g., ChatGPT, Bard). They are not finetuned to explain the buggy source code, and thus, their explanation could be generic or less insightful.
However, these LLMs are not exposed to structural information of the code (e.g., AST), neither they are fine-tuned to identify buggy code patterns, which might limit their capability of generating useful explanations.

In this paper, we propose \emph{Bugsplainer}\footnote{https://github.com/parvezmrobin/bugsplainer-webapp}, a novel web-based debugging solution that generates natural language explanations for software bugs by learning from a large corpus of bug-fix commits (i.e., commits that correct bugs).
Our solution leverages code structures to reason about buggy code and then generates appropriate explanations.
First, we leverage the code structures using an adaptation of structure-based traversal~\parencite{hu2018deep} to the buggy code.
Second, we train Bugsplainer using both buggy source code and its corrected version, which enables our tool to understand and detect buggy code patterns during its explanation generation for the buggy code.
Our tool is currently designed for Python programming language, but it can be easily adapted to other programming languages given its language-agnostic approach.
It provides the following features to explain software bugs to the developers.

\begin{enumerate}[label=(\alph*)]
    \item Given a code segment, Bugsplainer can generate succinct explanations for the bug in the code. 
    \item Provides three different variants of the explanation generation model -- Bugsplainer, Bugsplainer 220M, and Fine-tuned CodeT5, and can help developers understand bugs from different perspectives.
    \item Facilitates code changes on-the-fly and allows one to check how the changes in the code affect the generated explanations.
    \item Provides two working modes -- \emph{production} and \emph{experimental}, and allows a user to compare Bugsplainer with human written explanations and reason about the generated explanations.
\end{enumerate}

\begin{figure*}
    \centering
    \includegraphics[width=\linewidth]{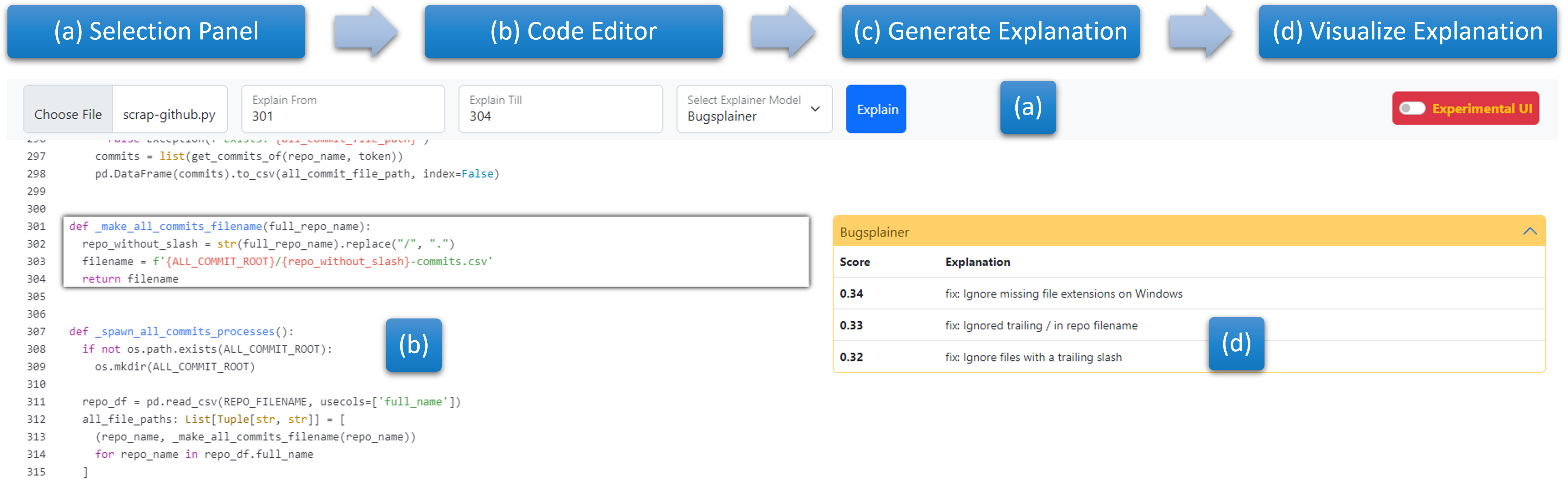}
    \caption{User interface of Bugsplainer}
    \label{fig: ui}
\end{figure*}

\section{Bugsplainer}
\label{sec: bugsplainer}
Fig.~\ref{fig: ui} shows the user interface of our tool -- Bugsplainer.
It is composed of four different components and has two different working modes. 
They are described as follows.

\subsection{Components}
\label{subsec: comp}
Bugsplainer is built using four different components -- (a) selection panel, (b) code editor, (c) explanation generator, and (d) explanation visualizer.
Among these four components, the explanation generator resides in a back-end server, whereas the remaining three reside in the front-end application.
In the following sections, we discuss these four components and their offered features.

\subsubsection{Selection Panel}
\label{sssec: selection-panel}

Fig.~\ref{fig: ui}a shows the selection panel -- the entry-point of Bugsplainer.
It contains a file selector which allows the user to choose a source code file from the local machine's file system.
Since at this stage Bugsplainer only supports python files (i.e., files with \texttt{.py} extension), the file selector filters out non-python files.
Upon selection, Bugsplainer reads the contents of the file and loads them into the code editor (see details in Section~\ref{sssec: code-editor}).

Besides the file selector, the selection panel contains two input fields to indicate the starting and ending line numbers of the code segment that needs an explanation.
It also contains a drop-down menu that allows one to select from three different models for explanation generation.
These models are designed using deep learning technology and are responsible for generating explanation against the given code segment.
The core difference among these models is in their learned parameters and how they analyze the buggy source code segment.


\subsubsection{Code Editor}
\label{sssec: code-editor}

Fig.~\ref{fig: ui}b shows the code editor of Bugsplainer.
Once a file is chosen from the selection panel, the code editor shows the contents of the file.
From there, the user can choose one or more lines of code for explanation.
This choice works synchronously with the selection panel.
For the user's convenience, the code editor highlights previously explained code segments as well as the current one.
Such highlighting not only prevents a user from generating duplicate explanations but also helps her quickly track back to previously explained code segments.

\subsubsection{Explanation Generator}
\label{sssec: exp-gen}

\begin{figure*}
    \centering
    \begin{subfigure}[b]{.20\linewidth}
        \begin{subfigure}[b]{\linewidth}
            \includegraphics[width=\linewidth]{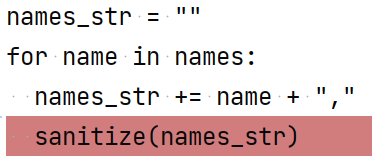}
            \caption{Buggy code}
            \label{subfig: example-buggy}
            \vspace{0.5em}
        \end{subfigure}
        \begin{subfigure}[b]{\linewidth}
            \includegraphics[width=\linewidth]{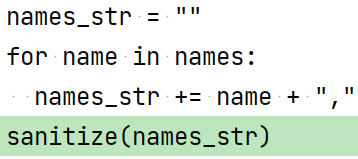}
            \caption{Bug-free code}
            \label{subfig: example-bug-free}
        \end{subfigure}
    \end{subfigure}
    \hspace{5mm}
    \begin{subfigure}[b]{.55\linewidth}
        \centering
        \includegraphics[width=\linewidth]{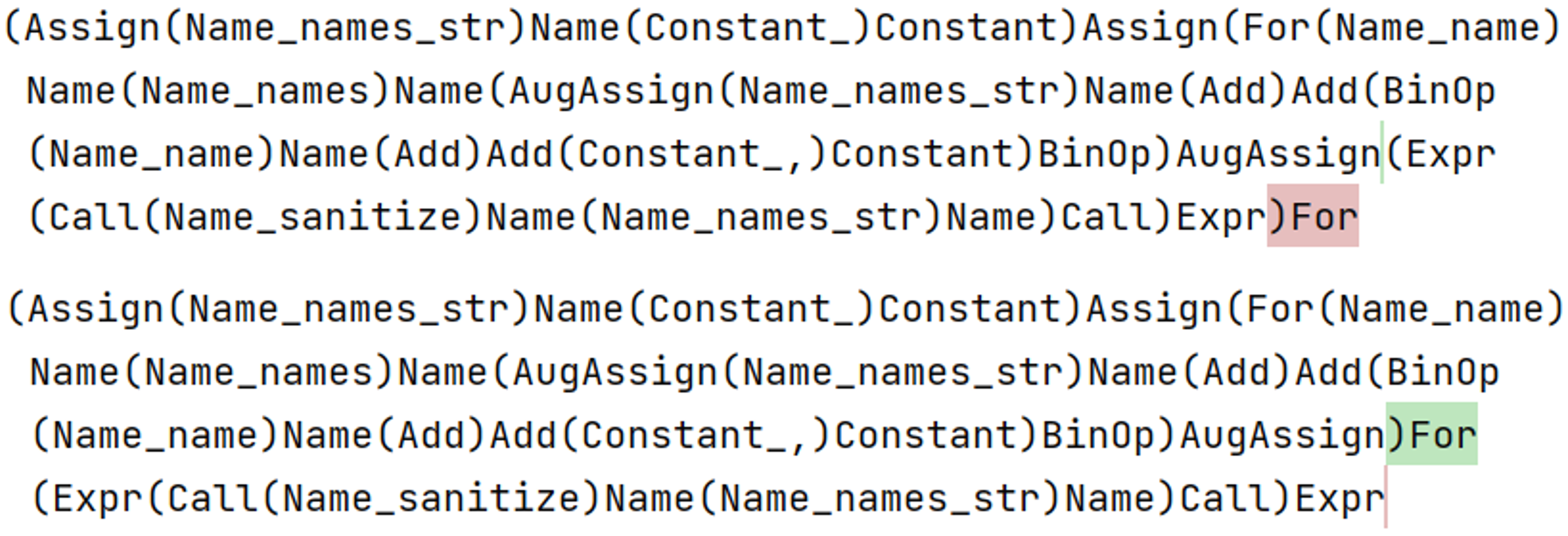}
        \caption{diffSBT sequence for the buggy and bug-free code}
        \label{subfig: example-diffsbt}
    \end{subfigure}

    \caption{An example of diffSBT sequence generation from buggy code and commit diff}
    \label{fig: diff-sbt}
\end{figure*}

In Bugsplainer, a back-end server has the core responsibility of generating the explanations.
Bugsplainer offers three different deep learning models for explanation generation -- Bugsplainer, Bugsplainer 220M and Fine-tuned CodeT5.
Among them, Bugsplainer and Bugsplainer 220M use structural information (e.g., abstract syntax tree) to generate the explanations.
Therefore, they can explain a bug even if it lies in the structure of the code.
On the contrary, Fine-tuned CodeT5 treats the source code as natural language text.
Depending on the context, these different deep learning models can focus on different aspects of the bug in generating their explanations, which can help the user understand a bug comprehensively.

Upon clicking the \emph{Explain} button in the selection panel (Section~\ref{sssec: selection-panel}), the front-end application sends an HTTP request to the back-end explanation generator.
The request contains the source code, starting and ending line numbers of the selected code segment, and the name of the selected deep learning model. 
Upon receiving them, the explanation generator generates the explanations as follows.

\textbf{Extract Buggy and Bug-free AST Nodes from Commit}:
First, it constructs abstract syntax trees (AST) from the source code document.
Using these line numbers, we extract the buggy nodes from the AST.
Besides the affected lines, the contextual information (e.g., surrounding lines) often provides useful clues about why the code was changed.
\textcite{asaduzzaman2014cscc} suggest that three lines of code around a target line might be sufficient to capture the contextual information.
While extracting the buggy and bug-free nodes, it thus also extracts the nodes representing three lines above and below the changed lines in the code.

\setlength{\textfloatsep}{2pt}
\begin{algorithm}[t]
    \small
    \caption{Generate diffSBT sequence from commit diff}
    \label{algo: diff-sbt}
    
    \begin{algorithmic}[1]
    \Function{diffSBT}{c}
        \Comment{Generate diffSBT sequence for commit}
        \State buggyAST $\gets$ \Call{BuildAST}{c.buggyCode}
        \State bugfreeAST $\gets$ \Call{BuildAst}{c.bugFreeCode}
        \State buggyNodes $\gets$ \Call{Intersections}{buggyAST, c.removed}
        \State bugfreeNodes $\gets$ \Call{Intersections}{bugfreeAST, c.added}
        \State \Return \Call{SBT}{buggyNodes} + $\langle /s\rangle$ + \Call{SBT}{bugfreeNodes}
    \EndFunction
    \Statex
    
    \Function{Intersections}{r, ln}
    
    \State nodes $\gets \phi$
    \Comment{Initialize nodes with an empty list}
    \ForAll{n in r}
    \Comment{Get intersections for all nodes in $r$}
    
        \If{\Call{IsInside}{n, ln} \textbf{or} \Call{IsExpression}{n}}
            \label{diffsbt: is-expression}
            \State \Call{Append}{nodes, n}
            
        \ElsIf{\Call{StartsInside}{n, ln}}
            \Comment{Keep $n$ but prune the children outside $ln$}
            \State n.children $\gets$ \Call{IntersectingChildren}{n, ln}
            \State \Call{Append}{nodes, n}
        
        \ElsIf{\Call{EndsInside}{n, ln}}
        \Comment{Node $n$ starts before the $ln$. Return only the children of $n$ that intersect with $ln$.}
            \State children $\gets$ \Call{IntersectingChildren}{n, ln}
            \State \Call{Append}{nodes, children}
        \EndIf
    
    \EndFor
    
    \State \Return nodes
    
    \EndFunction

    \Statex

    \Function{IntersectingChildren}{r, ln}
        \State children $\gets \phi$
        \ForAll{n in r.children}
            \If{\Call{HasIntersection}{n, ln}}
                \State node $\gets$ \Call{Intersections}{ch, ln}
                \State \Call{Append}{children, node}
            \EndIf
        \EndFor
        \State \Return children
    \EndFunction

    \end{algorithmic}

\end{algorithm}

\textbf{Generate diffSBT Sequence}:
In this step, it converts the AST nodes into diffSBT sequences using the diffSBT algorithm.
Algorithm~\ref{algo: diff-sbt} shows our algorithm -- \emph{diffSBT} -- for structure-preserving sequence generation from commit diff, which is an adaptation of SBT algorithm by \textcite{hu2018deep}.
diffSBT captures structural information from the partial AST by traversing the remaining nodes and edges. 
Fig.~\ref{fig: diff-sbt} shows two source code segments, where one of them is buggy (Fig.~\ref{subfig: example-buggy}), and the other is bug-free (Fig.~\ref{subfig: example-bug-free}).
They have exactly the same set of tokens, but the bug resides in their structures.
Fig.~\ref{subfig: example-diffsbt} shows the diffSBT sequence of these source code segments.
We see that the difference between them is clearly visible in the diffSBT sequence.
If the code is treated as natural language, it will be hard to detect this kind of bug.

\textbf{Generate Explanation}:
Once the diffSBT sequence is generated from the selected code segment, the deep-learning model generates explanations for the bug in the code utilizing its knowledge from pre-training and fine-tuning.
Our deep-learning models were pre-trained in two steps -- unsupervised pre-training and \emph{discriminatory pre-training}.
We reuse a CodeT5~\cite{wang2021codet5} model that was pre-trained with GitHub CodeSearchNet~\cite{husain2019codesearchnet} dataset.
Then, we performed our \emph{novel} discriminatory pre-training to the model, where it learns to generate explanation from both buggy and bug-free version of the code.
Along the way, it also learns to discriminate between buggy and bug-free code patterns.
Discriminatory pre-training has been shown to improve the final explanation generation by 4-35\%.
Finally, we fine-tuned our model to generate explanation from solely buggy code segment.

Utilizing the knowledge from the pre-training and fine-tuning phases, the model generates explanation for the given buggy code segment along with confidence score for each explanation.
Finally, the back-end server returns the top three explanations along with their confidence scores to the front-end application.

\subsubsection{Explanation Visualizer}
\label{sssec: viz-explain}
The Explanation Visualizer component (Fig.~\ref{fig: ui}d) is responsible for visualizing the explanations received from the back-end server in a user-friendly manner.
It groups all the received explanations based on the location of the source code segment.
Since there can be several explanations for the same code segment from different models, the Explanation Visualizer might get cluttered.
As a solution, we keep all the explanations collapsed by default.
When the user hovers on (i.e. moves the cursor on) an explanation group or corresponding source code segment, the explanations are shown along with their confidence score.

To differentiate between different explanation groups, we color the groups cyclically with six different colors.
This colors contrast with each other and the background according to the Web Content Accessibility Guidelines (WCAG)\footnote{https://www.w3.org/TR/WCAG}.
We also use a set colors that color-blind persons can distinguish~\parencite{Wong2011}.

\begin{figure*}
    \centering
    \includegraphics[width=.75\linewidth]{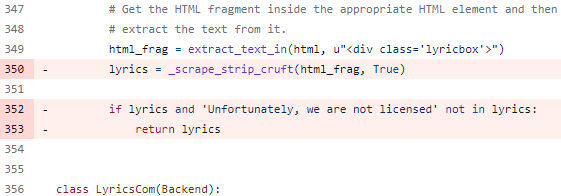}
    \caption{An example of buggy source code}
    \label{fig: motivate}
\end{figure*}

\subsection{Working Modes}
\label{subsec: modes}

Bugsplainer can work in two different modes -- \emph{production} and \emph{experimental}.
A user can choose either mode using a toggle button on the top right corner of the application (Fig.~\ref{fig: ui}).
The production mode is the default mode where Bugsplainer delivers automatically generated explanations for the buggy code as described in the above sections.

In the experimental mode, instead of selecting a file from the machine's file system, the user can select a file from a pre-defined set provided by Bugsplainer.
Each file from the set has a pre-defined bug location with human-written explanations associated with it.
Upon selection of a file, along with the file contents, Bugsplainer shows the corresponding human-written explanations. 
It also highlights the buggy code segments connected to these human-written explanations.
Then, the user can generate explanations for any code segment and compare them against the human-written explanations.
Such comparison not only helps the user grow confidence in Bugsplainer but also helps reason why Bugsplainer explains a bug in a certain way.

\section{A Use Case Scenario}
\label{sec: use-case}

\begin{figure*}
    \centering
    \includegraphics[width=\linewidth]{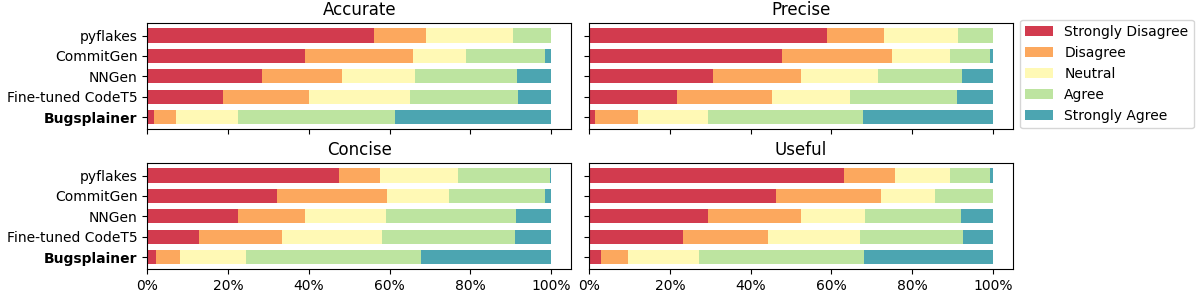}
    \caption{Comparison of Bugsplainer with the baselines using Likert scores}
    \label{fig: dev-study}
\end{figure*}

With a use case scenario, we attempt to explain how \emph{Bugsplainer} can help a software engineer in understanding a software bug.

Suppose a software developer, Alice, has been working on a software application that scrapes lyrics from an HTML page and shows the lyrics within the application.
Unfortunately, several users report that the application crashes during its use.
Alice is assigned to fix the bug.
She was able to find the location of the bug using a tool recently published.
That tool indicates that the bug resides between line 350 and line 353 in \texttt{beetsplug/lyrics.py} file, as shown in Fig.~\ref{fig: motivate}.
However, by just looking at the code, Alice could not understand what the problem was in the code.
Therefore, Alice decides to use a few static analysis tools to understand the problem in the code.
Unfortunately, the static analysis tool that Alice uses (e.g., pyflakes) could not find any issue in the code either.

Now, let us assume that Alice has decided to use \emph{Bugsplainer}.
As it is a web application, she does not need to install or configure anything on her local machine.
Bugsplainer is trained with thousands of bug-fixing changes.
It also can reason about the structural differences between buggy and bug-free code and leverages them to generate explanations for the buggy code (see details in Section~\ref{sssec: exp-gen}).
When Alice requests an explanation, Bugsplainer explains the above buggy code (Fig. \ref{fig: motivate}) as ``fix crash when lyrics not found."
From this explanation, Alice now understands the root cause of the bug.
That is, when the HTML page does not contain any lyrics, then the value of \texttt{html\_frag} is \texttt{None} (equivalent to \texttt{null} in many languages).
In such a case, any attempt to scrape from the HTML element will result in an application crash.
Therefore, Alice wraps the buggy lines (i.e., 350 -- 353) inside an \texttt{if} clause that checks whether \texttt{html\_frag} variable contains any value or not.
Such a code change solves the bug and prevents the application from crashing. Thus, as shown above, our tool -- Bugsplainer -- is able to provide an accurate and concise explanation that has helped Alice, the developer, better understand and fix the bug.

\section{Evaluation}
\label{sec: eval}

\looseness=-1
We evaluate Bugsplainer and its explanation generation models extensively.
The detailed evaluation results can be found in the corresponding technical paper~\cite{mahbub2022explaining}.
To evaluate the explanations generated by Bugsplainer, we use three different metrics -- BLEU score~\cite{papineni2002bleu}, Semantic Similarity~\cite{haque2022semantic}, and Exact Match.
Relevant studies~\cite{vaswani2017attention, raffel2020exploring, wu2016google, wang2021codet5} frequently used these metrics, which justifies our choice. 
To the best of our knowledge, there exists no work that explains software bugs in natural language texts. 
Since commit message generation is quite similar to explanation generation, we use state-of-the-art commit message generation techniques as our baseline.
The main difference between commit message generation and explanation generation is the former takes both buggy and bug-free lines as input whereas the latter takes only buggy lines as input.
In particular, we compare Bugsplainer with three commit message generation techniques namely -- \emph{CommitGen}~\cite{jiang2017automatically}, \emph{NNGen}~\cite{liu2018neural}, and \emph{Fine-tuned CodeT5}~\cite{wang2021codet5} and a static analysis tool \emph{pyflakes}~\cite{pyflakes}.
None of these existing approaches for commit message generation learns to differentiate between buggy and bug-free code. Thus, our approach has a better chance of generating meaningful explanations for the buggy code.

We find that Bugsplainer achieves a BLEU score of 33.15 and Bugsplainer 220M achieves 33.87, which are considered as \emph{understandable} and \emph{good} according to Google's AutoML Translation Documentation \cite{goole-automl}.
Among the baselines, Fine-tuned CodeT5 achieved the closest performance to Bugsplainer.
Nonetheless, Bugsplainer performance was better with a statistically significant margin i.e., \emph{p-value} = 0.008 $<$ 0.05, Cliff's \emph{d} = 1.0 (\emph{large}) for all three metrics.

To address the subjective nature of our research problem, we also conduct a developer study involving 20 professional developers to compare our tool with the baselines.
The developers come from six different countries with 1--10 years of programming experience and 1--7 years of bug-fixing experience.
The identity of each tool was kept hidden during the study to avoid any bias. 

\looseness=-1
Fig.~\ref{fig: dev-study} shows the distribution of participants' agreement levels in different aspects.
We see that the participants disagree with Bugsplainer very few times (e.g., highest 14\% times in precision) with a substantial level of agreement (e.g., highest 76\% in accuracy).
On the contrary, nearly half of the time the developers strongly disagree with pyflakes and CommitGen.
Such a disagreement with pyflakes is explainable since it does not generate any error message for 13 out of 15 cases.
However, CommitGen, even after explaining all cases, receives a high disagreement due to its generic and less informative explanations.
According to the developers, the closest competitor of Bugspaliner is Fine-tuned CodeT5.
We also perform the Wilcoxon Signed Rank test to check whether the developers' agreement with Bugsplainer is significantly higher than that of Fine-tuned CodeT5.
For accuracy, conciseness, precision and usefulness, the p-values were found as $5.16e{-23}$, $2.74e{-17}$, $7.45e{-18}$ and $2.63e{-23}$ respectively.
All of these p-values are below the threshold of $0.001$, which makes the difference significant.

\balance

\section{Threats to Validity}
Bugsplainer depends on bug localization tools for explanation generation, which could be considered as a source threat.
To mitigate this threat, Bugsplainer accepts a range of lines containing both buggy lines and their surrounding lines as input during explanation generation.
However, we do not indicate which lines among them contain a bug.
That is, it does not need to know the exact buggy lines for its operation.
Thus, precise localization of the bug either by developers or by existing tools might not be necessary to generate explanations using our tool.

\section{Conclusion}
\label{sec: conclusion}
Explaining a bug in software code is essential for understanding and fixing the bug. Unfortunately, not many studies aim to explain bugs hidden in the code to the developers. To address the problem, we propose a novel web-based debugging solution -- Bugsplainer -- that can generate natural language explanations for buggy code leveraging advanced deep learning technology (e.g., CodeT5). It allows a user to change the code on-the-fly and check how changes in the code affect the generated explanations.
Furthermore, in the experimental mode, it allows a user to compare generated explanations with human-written explanations, which is a key to growing user confidence in the application. In future, we plan to incorporate vulnerability prediction in our tool and extend our solution for other programming languages (e.g., Java, Javascript).

\vspace{1em}

\textbf{Acknowledgement:} This work was supported by Dalhousie University and Mitacs Accelerate International Program. We would like to thank \emph{Ben Reaves} and \emph{Massimiliano Genta} from our industry partner -- Metabob Inc. We would also like to thank all the anonymous participants in our developer study.

\scriptsize
\printbibliography
\end{document}